\documentclass[aps,reprint,amsmath,amssymb,floatfix]{revtex4-2}
\usepackage{graphicx}
\usepackage[colorlinks, citecolor=blue, linkcolor=blue]{hyperref}
\usepackage{txfonts}

\begin{document}

\title{Effects of delayed feedback on the power spectrum of spin-torque nano-oscillators}

\author{J{\'e}r{\^o}me Williame}
\author{Joo-Von Kim}
\email{joo-von.kim@c2n.upsaclay.fr}
\affiliation{Centre de Nanosciences et de Nanotechnologies, CNRS, Universit{\'e} Paris-Saclay, 91120 Palaiseau, France}

\date{16 Sep 2020}

\begin{abstract}
A theoretical study of delayed feedback in a spin-torque nano-oscillator model is presented. The feedback acts as a modulation of the supercriticality, which results in changes in the oscillator frequency through a strong nonlinearity, amplitude modulations, and a rich modulation sideband structure in the power spectrum at long delays. Modulation sidebands persist at finite temperatures but some of the complex structure is lost through the finite coherence time of the oscillations. 
\end{abstract}

\maketitle


\section{Introduction}

Spin-torque nano-oscillators (STNOs) represent nanoscale systems in which self-sustained magnetization oscillations are driven by electrical currents. STNOs typically comprise a magnetoresistive stack such as a spin valve or magnetic tunnel junction, whereby the traversal of spin-polarized currents can exert torques on the magnetization of the free magnetic layer~\cite{Berger:1996jd, Slonczewski:1996wo, Slonczewski:1999km}, giving rise to oscillations when the spin torques compensate on average the intrinsic Gilbert damping over a period of oscillation. The oscillations can involve a wide range of dynamical modes, such as large angle precession of spin wave eigenmodes~\cite{Kiselev:2003hp, Krivorotov:2005fx, Mistral:2006er, Houssameddine:2007gs, Cornelissen:2009id, Zeng:2012bf}, vortex gyration~\cite{Ivanov:2007kn, Pribiag:2007dk, Pufall:2007jc, Mistral:2008js, Dussaux:2010ef, Khvalkovskiy:2010if, Locatelli:2011hw}, skyrmion motion~\cite{GarciaSanchez:2016cx}, and dynamical droplet solitons~\cite{Hoefer:2010ka, Mohseni:2013eh}. Similar phenomena can be observed with the spin Hall effect, where spin currents generated in an adjacent layer with strong spin orbit coupling exerts torques on the magnetization~\cite{Demidov:2012im, Smith:2014et, Awad:2016eo, Divinskiy:2017ba, Evelt:2018bs, Dvornik:2018ei, Spicer:2018gd, Hamid:2018gz}. This geometry is particularly attractive for delayed feedback effects, where the oscillator output is re-injected into the input after amplification and a time delay. The spin Hall effect allows for a three-terminal device in which the drive and readout currents can be decoupled~\cite{Liu:2012gp, Jue:2018dm, Jue:2018gj}.

The presence of thermal noise results in a finite coherence time for the oscillations, so it is important to understand and control this aspect for potential applications. Because of the strong frequency nonlinearity that is a characteristic of STNOs, strong coupling between the phase and amplitude fluctuations leads to a large contribution to inhomogeneous broadening in such systems~\cite{Kim:2008ic, Kim:2008gs, Tiberkevich:2008jh, Bianchini:2010eq, Quinsat:2010kt, Taniguchi:2014eh}. Since thermal fluctuations are unavoidable at ambient temperatures, external means can be used to mitigate spectral line broadening. One approach is through phase-locked loops~\cite{Keller:2009ix}. It has been shown that self-injection locking in a vortex-based STNO can be efficient for reducing the spectral linewidth~\cite{Khalsa:2015kn, Tsunegi:2016ka}. It has also been demonstrated that an STNO subjected to delayed feedback can improve its sensitivity to external signals~\cite{Tiberkevich:2014eh}.

Other aspects of feedback in spintronic systems have also been explored. For example, it has been shown experimentally that feedback through applied magnetic fields, where the time-varying output of the magnetoresistance from a magnetic tunnel junction, for example, is re-injected into a co-planar waveguide to generate a microwave field. This can lead to self-oscillations~\cite{Kumar:2016io} and self-injection locking~\cite{Singh:2018gr}. Such feedback loops lead to the appearance of complex power spectra with several modulation sidebands~\cite{Singh:2017gt}. A similar phenomenon can also be achieved using the spin Hall effect as a feedback mechanism~\cite{Bhuktare:2017ja}. Feedback in spin-torque nano-oscillators can also result in chaotic phenomena~\cite{Williame:2019hd, Taniguchi:2019ej}, where the time delay increases the dimensionality of the phase space thereby allowing chaos to appear even for nominally low-dimensional systems. The memory depth of a time-delay architecture for reservoir computing has also been explored in vortex-based systems~\cite{Yamaguchi:2020hg}.

In this article, we present a theoretical study of the role of delayed feedback in a model of a spin-torque nano-oscillator. The article is organised as follows. In Section II, we briefly describe the oscillator model and the implementation of the feedback. In Section III, we examine the role of delayed feedback at zero temperature. The effect of feedback at finite temperatures is discussed in Section IV, where particular attention is given to the variation of the spectral power and linewidth. Finally, a discussion is given in Section V and concluding remarks are given in Section VI.

\section{Model}
We consider the following model of a spin-torque nano-oscillator, which can be derived from spin wave theory~\cite{Slavin:2005cl, Rezende:2005hd, Slavin:2009fx, Kim:2012du} and describes a limit cycle oscillator with a strong frequency nonlinearity,
\begin{equation}
\frac{dc}{dt} = -i \left( \Omega_0 + N |c|^2 \right) c - \left[ \Gamma - \sigma I \left( 1 -  |c|^2 \right) \right] c.
\label{eq:nom}  
\end{equation}
$c$ is a dimensionless complex variable that describes the amplitude and phase of the oscillator. $\Omega_0$ is the linear frequency, $N$ is the nonlinear frequency parameter, $\Gamma$ represents linear viscous damping, and $\sigma I$ represents the spin-torque that plays the role of negative damping for positive currents, $I>0$. We choose the linear damping rate to define a natural time scale, $t = \Gamma t'$, which allows us to express the dynamics with the dimensionless time $t'$,
\begin{equation}
\frac{dc}{dt'} = -i \left( \omega_0 + n |c|^2 \right) c - \left[1 - \zeta \left( 1 - |c|^2 \right)  \right] c,  
\end{equation}
where $\omega_0 = \Omega_0/\Gamma$ is the reduced frequency, $n = N/\Gamma$ is the reduced nonlinear frequency shift, and $\zeta = \sigma I / \Gamma$ is the supercriticality parameter. We assume that the oscillator signal can be read out from the real part of $c$, which is subsequently used as an additional feedback signal into the drive current. Physically, this might represent one component of the magnetization vector such as $m_x$, where $x$ would be the direction of the reference layer magnetization that defines the amplitude of a magnetoresistance variation~\cite{Kim:2006im}. With $G$ representing the feedback gain and $\tau$ the delay, we write
\begin{multline}
\frac{dc}{dt'} = -i \left( \omega_0 + n |c|^2 \right) c \\
- \left[1 - \zeta \left( 1 + G \, \mathrm{Re}\left[c(t'-\tau)   \right] \right) \left( 1 - |c|^2 \right)  \right] c,  
\end{multline}
We can simplify the equations of motion further by transforming into the rotating frame of the oscillator based on the linear frequency, $c(t') = c(t') \exp(-i \omega_0 t')$. This results in 
\begin{equation}
\frac{dc}{dt'} = -i n |c|^2 c - c + \zeta \left[ 1 + G g(t'-\tau)\right] \left(1-|c|^2 \right) c,
\label{eq:EOM}
\end{equation}
where $g$ represents the feedback function
\begin{equation}
g(t'-\tau) = \mathrm{Re}\left( c(t'-\tau)  \exp\left[-i \omega_0 (t'-\tau) \right] \right).
\end{equation}
Since $c(t')$ now represents the dynamics in the rotating frame, its frequency is governed primarily by the nonlinear frequency term $n$. As such, the feedback term contains a product of two frequencies: the linear frequency $\omega_0$ and the nonlinear shift $n$. Moreover, the feedback enters as a parametric excitation, since it modulates the supercriticality parameter $\zeta$.

\section{Feedback dynamics at zero temperature}
We first examine the role of delayed feedback in the absence of thermal fluctuations. We perform a time-integration of the dynamics in Eq.~(\ref{eq:EOM}) in the supercritical region, $\zeta = 2$, which is far above threshold $(\zeta = 1)$ for self-sustained oscillations. Because Eq.~(\ref{eq:EOM}) describes the dynamics in the rotating frame, the solution is given by a fixed point in phase space, $c(t') = c_0 = \sqrt{\left(\zeta-1\right)/\zeta}$, in the absence of frequency nonlinearity and delayed feedback, $n = G = 0$. This describes the dynamics of a phase oscillator. When $n \neq 0$, on the other hand, the dynamics of $c(t')$ is described by a circular limit cycle in phase space with a radius of $c_0$ and phase velocity of $n (\zeta -1)/\zeta$.

\begin{figure}
\centering\includegraphics[width=8.5cm]{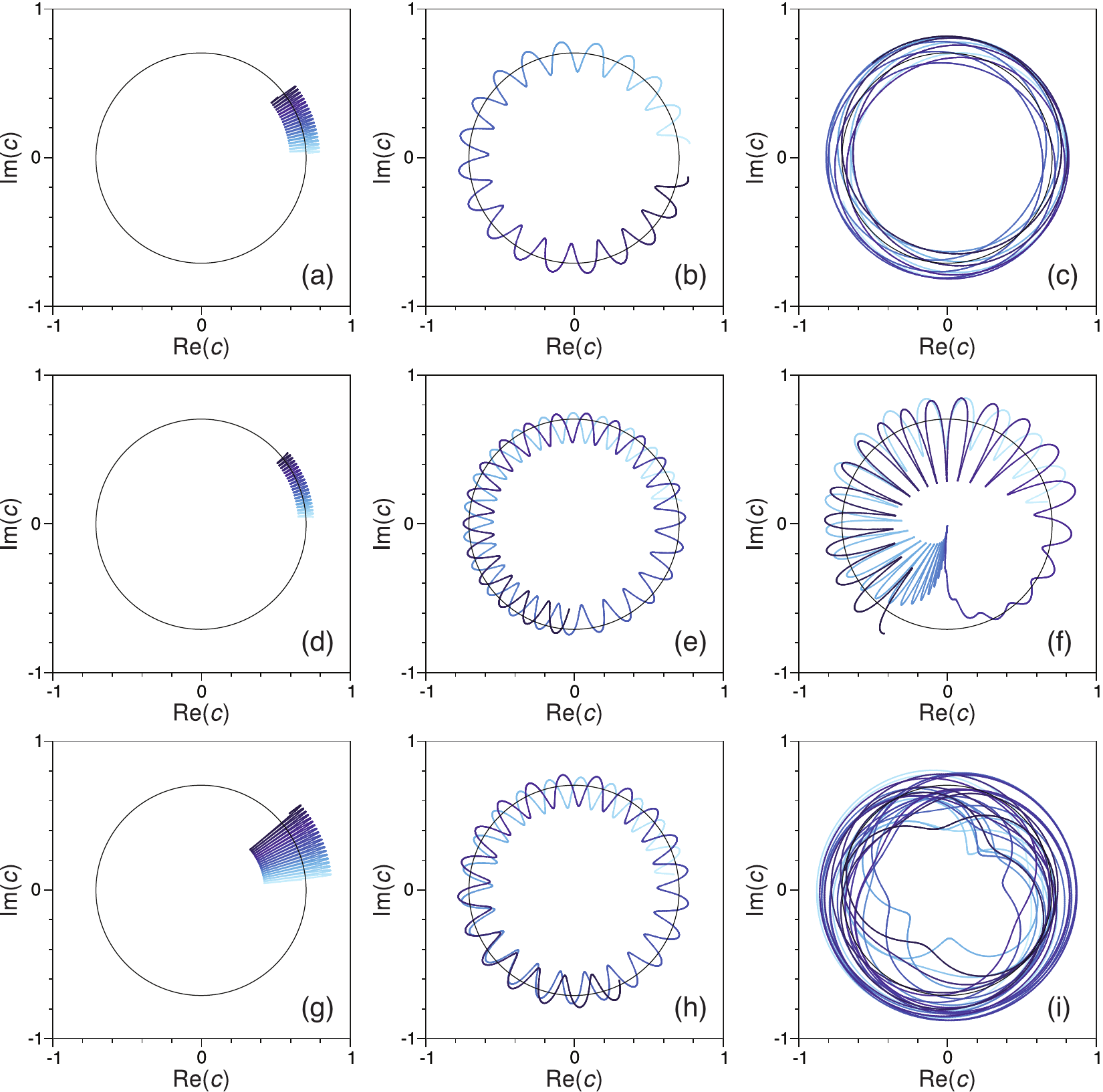}
\caption{Phase space dynamics for different values of the frequency nonlinearity $n$, feedback gain $G$, and feedback delay $\tau$. (a) $n = 0.1$, $G = 2$, $\tau = 1$. (b) $n = 1$, $G = 2$, $\tau = 1$. (c) $n = 10$, $G = 2$, $\tau = 1$. (d) $n = 0.1$, $G = 1$, $\tau = 1$. (e) $n = 1$, $G = 2$, $\tau = 10$. (f) $n = 1$, $G = 5$, $\tau = 200$. (g) $n = 0.1$, $G = 5$, $\tau = 1$. (h) $n = 1$, $G = 2$, $\tau = 100$. (i) $n = 10$, $G = 2$, $\tau = 200$ The trajectory is shown for an interval of $\Delta t' = 4\pi$ in (a,b,c,d,g), $\Delta t' = 8\pi$ in (e,h), and $\Delta t' = 12\pi$ in (f,i), which correspond to 20, 40, and 60 periods of the linear oscillator ($\Omega = 10$), respectively. The circle of radius $c_0 = 1/\sqrt{2}$ represents the limit cycle in the absence of delayed feedback. The colour code represents time evolution, where light to dark blue corresponds to increasing $t'$.}
\label{fig:phase}
\end{figure}

Let us now consider the role of delayed feedback, in particular the effect of the frequency nonlinearity $n$, gain $G$, and delay $\tau$ in the amplitude dynamics. Some results of the time-integration of Eq.~(\ref{eq:EOM}) are given in Fig.~\ref{fig:phase}.
Here, we plot the evolution of $c$ in the rotating frame over a small time interval to highlight its dynamics, where a colour code is used to indicate the arrow of time. The effect of the frequency nonlinearity can be seen in Figs.~\ref{fig:phase}(a)-(c). The feedback leads to modulations of the oscillator amplitude at a rate given by the linear frequency, $\Omega_0$, but how quickly this modulation propagates in the rotating frame depends on the amplitude of $n$. At weak nonlinearity [$n=0.1$, Fig.~\ref{fig:phase}(a)], the feedback-induced modulation of the oscillator amplitude about $c_0$ drifts with a low angular velocity, resulting in a sinusoidal pattern that remains close to the initial point. As the nonlinearity is increased to ($n = 1$), the angular velocity increases accordingly and we can clearly observe the modulation across nearly the full circle, as shown in Fig.~\ref{fig:phase}(b). As the angular velocity is further increased with the frequency nonlinearity ($n=10$), the feedback-induced modulation now occurs at a rate that is slower than the angular velocity in the rotating frame, so the modulation now appears like a ringed structure as shown in Fig.~\ref{fig:phase}(c). The overall effect is a strong modulation in the oscillator amplitude that also translates into modulations in the angular frequency as a result of the $n$ term. The role of the feedback gain $G$ can be seen in Figs.~\ref{fig:phase}(a), \ref{fig:phase}(d), and \ref{fig:phase}(g). The main effect is the magnitude of the amplitude modulation, which increases with increasing $G$.

The influence of the time delay $\tau$ on the phase space dynamics can be seen in Figs.~\ref{fig:phase}(b), \ref{fig:phase}(e), and \ref{fig:phase}(h), where the delay is varied from $\tau = 1$, $\tau = 10$, to $\tau = 100$, respectively. We can ascertain a change in the relative phases of the modulation as $c$ cycles around the unit circle. More drastic effects can occur at long delays ($\tau = 200$). For a large gain, $G = 5$ for example, we can observe instances of temporary oscillator death, which is represented by the fast spiral toward $c=0$ in Fig.~\ref{fig:phase}(f), which recovers to an oscillatory state shortly afterwards. The presence of a strong nonlinearity ($n = 10$) can result in aperiodic motion, which can be seen in Fig.~\ref{fig:phase}(i). Both of these scenario are suggestive of chaotic dynamics.

\begin{figure}
\centering\includegraphics[width=8cm]{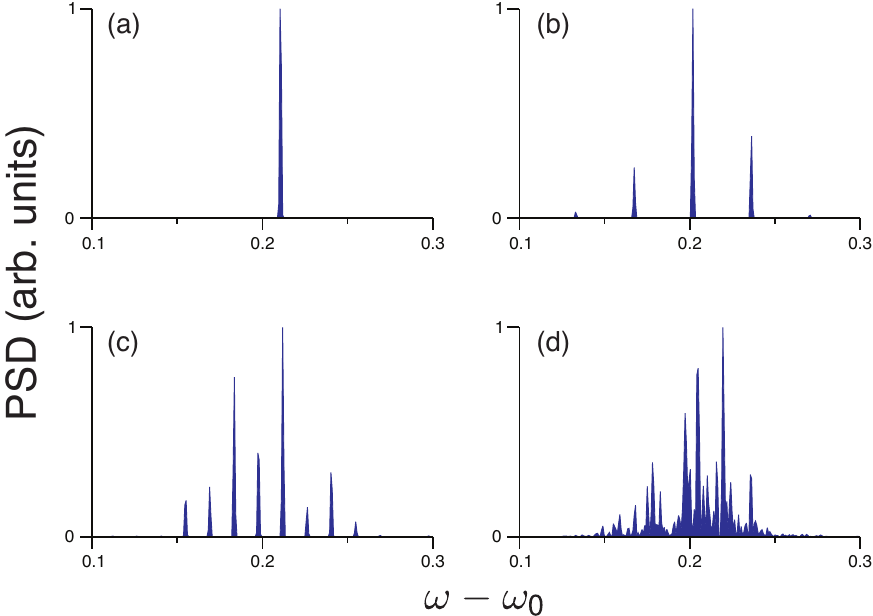}
\caption{Power spectra of $c$ in the rotating frame with $\zeta = 1.25$, $G = 2$, $n =1$, and different delays: (a) $\tau = 1$, (b) $\tau = 60$, (c) $\tau = 106$, (d) $\tau = 200$. }
\label{fig:spectra_LF}
\end{figure}

The role of the feedback delay on the power spectrum of $c$ is shown in Fig.~\ref{fig:spectra_LF} at different delays with $\zeta=1.25$, $G=2$, and $n=1$.
The power spectra are computed as follows. For a given set of parameters, time integration of Eq.~\ref{eq:EOM} is performed using the Euler-Maruyama method with a time step of $\Delta t' = 2^{-6}$ up to the maximum time of $t_\mathrm{max}' = 2^{17}-\Delta t'$, resulting in a time series comprising $2^{23}$ points. The Welch method~\cite{Welch:1967} is then applied to compute the resulting power spectrum, which involves: (i) decomposing the time series data into half-overlapping windows comprising $2^{18}$ points; (ii) applying a Hann filter to the windowed data to minimise spectral leakage; (iii) performing a Fourier transform of this filtered, windowed data; and finally (iv) averaging over the Fourier transforms to obtain the power spectrum. At short delays, the power spectrum is dominated by a peak near $\omega - \Omega \approx 0.2$ [Fig.~\ref{fig:spectra_LF}(a)], which is similar to the dynamics without feedback where the position of the peak is given by the $n|c|^2$ term. As the delay is increased, we observe a clear signature of modulation with the appearance of well-defined sidebands around the central frequency peak [Fig.~\ref{fig:spectra_LF}(b)]. The number of sidebands increases as the delay increases [Fig.~\ref{fig:spectra_LF}(c)], resulting in a broad spectral line at long delays [Fig.~\ref{fig:spectra_LF}(d)].

\begin{figure}
\centering\includegraphics[width=8.5cm]{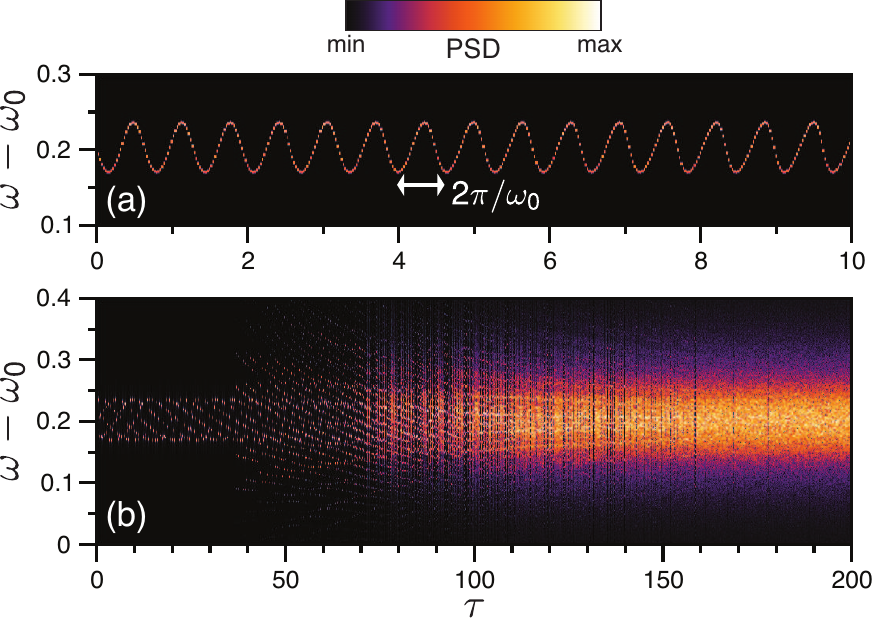}
\caption{Colour map of the power spectral density as a function of delay $\tau$ at $\zeta = 1.25$ for $n=1$ and $G=2$. (a) Short delays. (b) Long delays. The oscillation period in (a), $2\pi/\Omega$, is related to the linear frequency $\omega_0$ that enters the feedback term.}
\label{fig:PSD_map_LF}
\end{figure}

The different modulation regimes can more clearly be seen in Fig.~\ref{fig:PSD_map_LF}, where a colour map of the power spectral density is presented as a function of $\tau$.
The colour map corresponds to the power spectral density presented in Fig.~\ref{fig:spectra_LF}. At short delays, the power spectrum is dominated by a single peak close to $\omega - \omega_0 \approx 0.2$, as shown in Fig.~\ref{fig:spectra_LF}(a), with a frequency that oscillates as a function of $\tau$. The period of this oscillation is given by $2\pi/\omega_0$, which corresponds to the periodic forcing in the feedback due to the linear frequency term.  This variation in the oscillator frequency is made possible by the nonlinearity $n$, since modulations in the amplitude $|c|$ through the feedback is translated to a time-dependent frequency shift. A similar phenomenon has been observed in vortex oscillators~\cite{Tsunegi:2016ka}, where self-injection has been shown to produce an oscillatory dependence of the oscillator frequency on the time delay. As the delay is further increased, additional modulation side bands appear as shown in Figs.~\ref{fig:spectra_LF}(b) and \ref{fig:spectra_LF}(c). These bands appear gradually from about $\tau \approx 40$ onwards and become closer to the central frequency peak. For $\tau \gtrsim 70$, the modulation sidebands become more numerous and indistinguishable from one another, resulting in broad features in the power spectrum like the example shown in Fig.~\ref{fig:spectra_LF}(d). This trend persists as the delay is increased to the maximum value considered $\tau = 200$, with the exception of certain values of the delay at which synchronization where well-defined sidebands can be seen. These points are characterized by the thin dark lines on Fig.~\ref{fig:PSD_map_LF}, which indicate an absence of broadening and a power spectrum that resembles the case shown in Fig.~\ref{fig:spectra_LF}(c).

\section{Feedback dynamics at finite temperature}
To account for the effects of finite temperatures, we supplement the equation of motion (\ref{eq:EOM}) with a stochastic term $f$,
\begin{equation}
\frac{dc}{dt'} = -i n |c|^2 c - c + \zeta \left[ 1 + G g(t'-\tau)\right] \left(1-|c|^2 \right) c + f(t').
\label{eq:SEOM}
\end{equation}
$f$ is assumed to represent Gaussian white noise with zero mean, $\langle f(t) \rangle =0$, and the spectral properties
\begin{equation}
\langle f^*(t_1)f(t_2) \rangle = 2 q \, \delta(t_1 - t_2), 
\end{equation}
where $q$ represents the amplitude of the thermal noise and is proportional to the temperature. With the fluctuation-dissipation theorem, it can be shown that the noise amplitude $q$ can be related to the phenomenological parameters associated with magnetization dynamics in the micromagnetic description~\cite{Kim:2012du}, which in the reduced units is given by
\begin{equation}
q = \frac{k_B T}{\hbar \Omega_0} \frac{g \mu_B}{M_s V},
\end{equation}
where $g$ is the electron g-factor, $\mu_B$ is the Bohr magneton, $k_B$ is Boltzmann's constant, $T$ is the temperature, $M_s$ is the saturation magnetization, and $V$ is the volume of the excitation. It is implicitly assumed that the spectral properties of $f$ remains unchanged between the laboratory and rotating frames.

\begin{figure}
\centering\includegraphics[width=8cm]{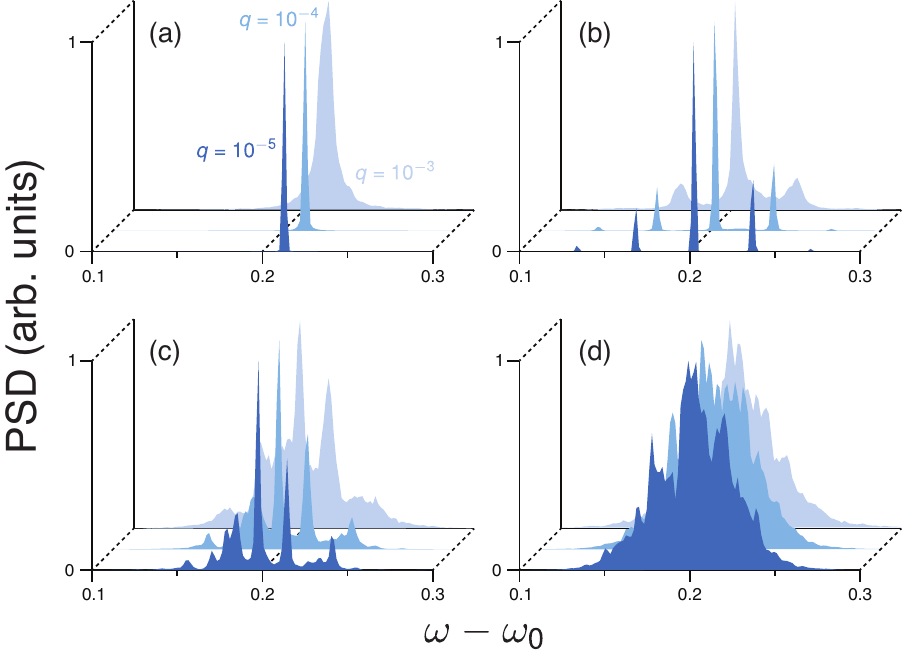}
\caption{Power spectra of $c$ in the rotating frame with $\zeta = 1.25$, $G = 2$, $n =1$, and different temperatures $q$ and delays: (a) $\tau = 1$, (b) $\tau = 60$, (c) $\tau = 106$, (d) $\tau = 200$.}
\label{fig:spectra_LF_q}
\end{figure}

The effect of thermal noise on the power spectra for different delays is shown in Fig.~~\ref{fig:spectra_LF_q}.
At short delays at which no modulation is apparent [Fig.~\ref{fig:spectra_LF_q}(a)], the noise results in a broadening of the main spectral line as expected from stochastic oscillator theory. We have verified that our numerical integration reproduces the main features of this theory, namely the  variation of the linewidth as a function of supercriticality far below and above threshold. At the onset of modulation, we observe that the modulation sidebands persist but experience broadening like the central peak [Fig.~\ref{fig:spectra_LF_q}(b)]. As the delay is increased, however, the number of sidebands present in Figs.~\ref{fig:spectra_LF}(c) and \ref{fig:spectra_LF}(d) are reduced or smeared out as a result of the line broadening and loss in temporal coherence due to the thermal noise. This particularly evident for $\tau = 200$ [Fig.~\ref{fig:spectra_LF_q}(c)], where the forest of peaks in Fig.~\ref{fig:spectra_LF}(d) evolves into a broad distribution in the power spectrum as the temperature is increased. This suggest that the coherence of the delayed feedback signal is lost and the overall result is akin to an additional noise term driving the oscillator dynamics.

\begin{figure}
\centering\includegraphics[width=8.5cm]{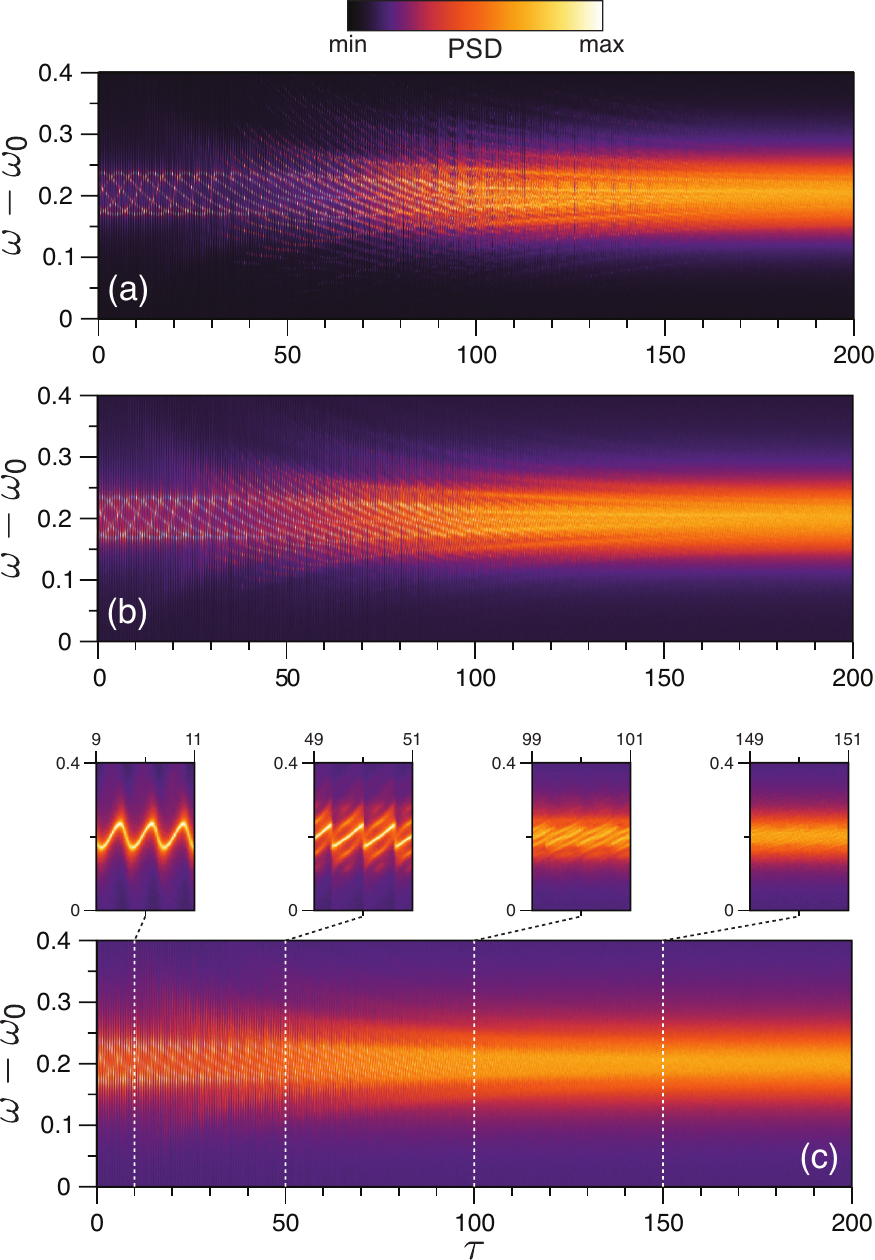}
\caption{Colour map of the power spectral density as a function of delay $\tau$ ($\zeta = 1.25$, $n=1$, $G=2$) at three different temperatures: (a) $q = 10^{-5}$, (b) $q = 10^{-4}$, and (c) $q = 10^{-3}$. The insets in (c) represent detailed views at four different intervals of the delay.}
\label{fig:PSD_map_LF_q}
\end{figure}

Figure~\ref{fig:PSD_map_LF_q} shows the colour maps of the power spectral density as a function of delay at three different temperatures.
Qualitatively, the overall variation of the spectral features with the feedback delay remains unchanged with respect to the behaviour seen in Fig.~\ref{fig:PSD_map_LF}. The primary effect of the thermal noise is to broaden the central and modulation peaks, which leads to a smearing out of the power spectra at large delays at which a multitude of peaks is seen at zero temperature. This phenomenon can be seen as the noise level is increased successively by an order of magnitude in Fig.~\ref{fig:PSD_map_LF_q}, where the position at which the modulation sidebands become visually indistinguishable from the broad central peak moves toward smaller values of $\tau$ as $q$ is increased. This loss in coherence can also be seen in the insets of Fig.~\ref{fig:PSD_map_LF_q}(c), where a detailed view of the power spectral density is given for four different intervals of the delay. The variation in the frequency of the central peak is accompanied by line broadening for $\tau \approx 10$, while a similar variation with modulation sidebands can be observed for delays up to $\tau \approx 50$. At around $\tau \approx 100$ the modulation is still visible but the sidebands barely perceptible; these features gradually vanish at larger delays ($\tau \approx 150$).

\begin{figure}
\centering\includegraphics[width=8.0cm]{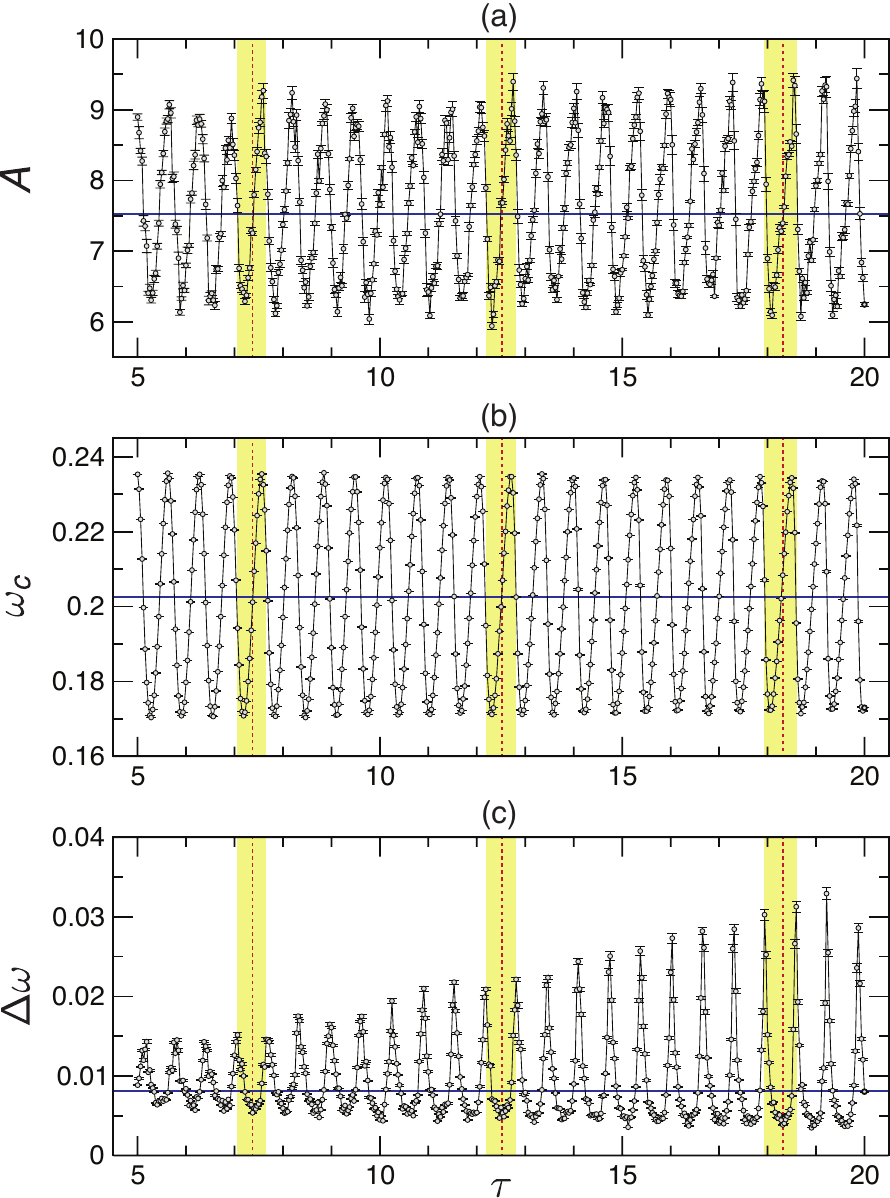}
\caption{Variation of the spectral line characteristics as a function of delay for $q=10^{-3}$. (a) Power, (b) frequency, (c) fullwidth at half maximum. Errors bars indicate errors in the parameters obtained from fits to a Lorentzian line shape. The blue horizontal lines indicate the values in the absence of delayed feedback. The dashed vertical lines indicate certain values at which $\Delta \omega$ is a minimum. The shaded yellow regions indicate an interval between two successive maxima in $\Delta \omega$.}
\label{fig:PSD_params}
\end{figure}

The power spectrum is dominated by the central frequency peak at short delays. For value of the frequency nonlinearity used ($n=1$) and the temperature range considered, the line shape is well described by a Lorentzian profile,
\begin{equation}
S(\omega) = \frac{1}{\pi} \frac{A \Delta \omega}{(\omega-\omega_c)^2+ (\Delta \omega/2)^2},
\end{equation}
where $\omega_c$ is the central frequency, $\Delta\omega$ is the full width at half maximum, and $A$ is the power. We have verified that $\Delta\omega$ varies linearly as a function of $q$ over the temperatures considered, which is consistent with behaviour resulting in a Lorentzian line shape as predicted by stochastic oscillator theory~\cite{Tiberkevich:2008jh}. (A Gaussian line shape is associated with a linewidth that varies like $\sqrt{q}$, which occurs when inhomogeneous broadening due to phase-amplitude coupling is important). The results of the fits to the spectral lines at short delays at $q=0.001$ are shown in Fig.~\ref{fig:PSD_params}.
As with the central frequency, we can observe oscillations in the oscillator power with the feedback delay [Fig.~\ref{fig:PSD_params}(a)]. The oscillations of the central frequency $\omega_c$ can be seen in Fig.~\ref{fig:PSD_params}(b). The variations take on a sinusoidal form at short delays and progressively evolve toward a sawtooth form as $\tau$ is increased. This can be understood as follows. At short delays, the feedback acts as a simple modulation of the oscillator amplitude, which through the frequency nonlinearity results in a modulation of the central frequency. As the delay increases the modulation acts more like a phase-locking mechanism, where the sawtooth form is a result of frequency pulling. We note that the variations in the frequency are centred about the value without delayed feedback, which is indicated by the horizontal blue line in Fig.~\ref{fig:PSD_params}(b).

The linewidth variations with the delay are shown in Fig.~\ref{fig:PSD_params}(c). In contrast to the oscillations in the power and frequency, the linewidth variations are not centred about the value without feedback. Minima in the $\Delta \omega$ occur when $\omega_c$ attains its value without feedback, as shown for three values of $\tau$ by the vertical dashed red lines in Fig.~\ref{fig:PSD_params}(c). This is consistent with a phase-locking phenomenon, where a similar reduction in the linewidth appears when an oscillator is phase-locked to an external source. Similar behaviour has been seen experimentally for vortex-based spin-torque nano-oscillators~\cite{Tsunegi:2016ka}. When $\omega_c$ deviates from its value without feedback, we observe a significant increase in the linewidth, where the value of the maxima appear to increase with the delay. The yellow shaded regions in Fig.~\ref{fig:PSD_params}(c) denote intervals between two successive maxima in the linewidth, which correspond to extrema in $\omega_c$. This indicates that the loss in phase coherence is maximum when the central frequency deviates the most from its value without feedback.

\begin{figure}
\centering\includegraphics[width=8.5cm]{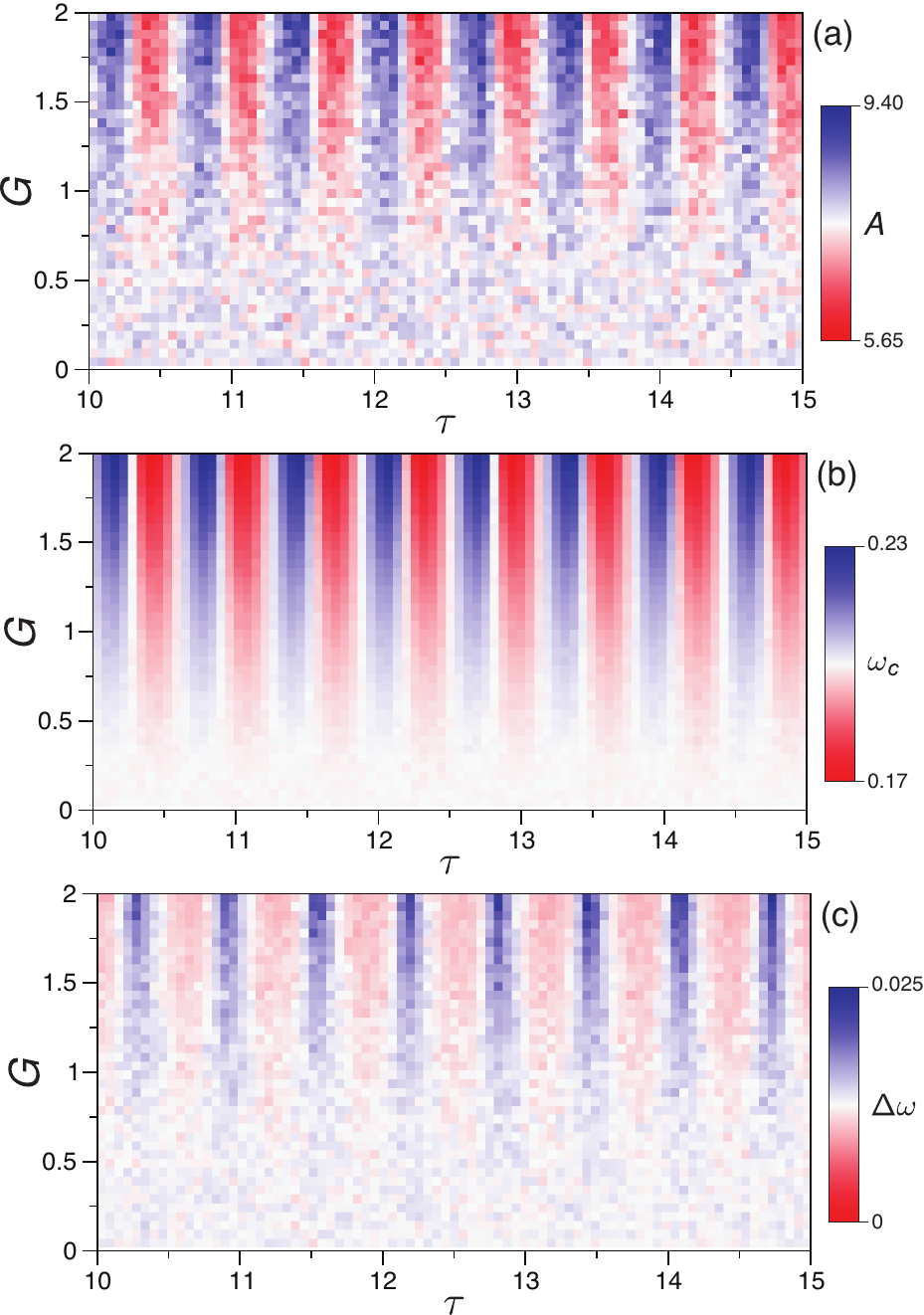}
\caption{Colour maps representing the variation of the spectral line characteristics as a function of delay $\tau$ and gain $G$ for $q=10^{-3}$. (a) Power, (b) frequency, (c) full width at half maximum. The colour scheme reflects the deviations with respect to the values without feedback.}
\label{fig:Params_Gtau}
\end{figure}

The variation of the spectral line properties with feedback delay and gain is presented in Fig.~\ref{fig:Params_Gtau}.
We observe that the overall qualitative behaviour illustrated in Fig.~\ref{fig:PSD_params}, i.e., the oscillations in the power, frequency, and linewidth with delay, is present all at values of the gain. The main difference lies in the magnitude of these variations with respect to the value without feedback, which is reflected in the colour scheme used where light grey indicates this value.

\section{Discussion}
We note that there are strong similarities but also key differences between our results and previous work, notably on delayed-feedback in vortex nano-oscillators~\cite{Khalsa:2015kn, Tsunegi:2016ka}. In the absence of feedback, it has been shown that the Thiele equation describing the vortex dynamics can be cast in the form of Eq.~\ref{eq:nom} through suitable transformations of the vortex variables~\cite{Grimaldi:2014iz}. However, the presence of in-plane field-like and damping-like spin-torques for the vortex oscillator, which do not contribute to the supercriticality, can have a profound influence on self-injection locking. Such features are not accounted for in the model discussed here, but could be incorporated through the inclusion of other forcing terms, such as those considered in earlier work on mutual synchronization of spin-torque nano-oscillators based on the formalism in Eq.~\ref{eq:nom}~\cite{Slavin:2005ga}.

\section{Conclusion}
In summary, we have presented a theoretical study on the effects of delayed feedback on the power spectra of spin-torque nano-oscillators. We have considered a nonlinear oscillator model derived from spin wave theory in which a single mode dominates the dynamics. Feedback has been introduced as a modulation of the drive current where the real part of the oscillator variable is taken to be the feedback signal. The amplitude and phase dynamics of the oscillator has been studied as a function of the feedback delay and gain.

At zero temperature, the feedback results in a strong modulation of the oscillator signal that gives rise to a large number of modulation sidebands in the power spectrum as the delay increases. Initially, the delay causes an oscillation in the central frequency as a function of delay as a result of the frequency nonlinearity. This persists with the appearance of the modulation sidebands, which eventually lead to a large broadening of the main peak when the density of the sidebands become large at long delays.

At finite temperatures, thermal noise leads to a broadening of the main spectral line and the modulation sidebands, as expected. The reduction in the coherence time due to thermal noise also limits the number of modulation sidebands visible, where some of the fine structure seen at zero temperature is smeared out. At short delays, Lorentzian fits to the spectral line reveal large oscillations in the linewidth, with minima coinciding with delays at which the oscillator frequency coincides with its value without feedback and maxima coinciding at values where the frequency deviation is largest.


\begin{acknowledgments}
This work was supported by the Agence Nationale de la Recherche (France) under contract nos. ANR-14-CE26-0021 (Memos) and ANR-17-CE24-0008 (CHIPMuNCS).
\end{acknowledgments}

\bibliographystyle{iopart-num}
\bibliography{articles}

\end{document}